\documentclass{optica-article}

\journal{opticajournal} 

\articletype{Research Article}

\usepackage{lineno}
\usepackage{graphicx}
\usepackage{dcolumn}
\usepackage{bm}

\usepackage{amsmath}
\usepackage{amsfonts} 
\usepackage{braket}
\usepackage{booktabs}
\usepackage{subfigure}
\usepackage{float}
\usepackage{siunitx}
\usepackage{float}
\usepackage{stfloats}

\begin{document}

\title{Experimental demonstration of improved reference-frame-independent quantum key distribution over 175km}

\author{Zhiyu Tian \authormark{1}, Ziran Xie \authormark{1}, Rong Wang \authormark{2}, Chunmei Zhang \authormark{3}, Shihai Sun  \authormark{1,*}}

\date{\today}

\address{\authormark{1}School of Electronics and Communication Engineering, Sun Yat-sen University, Shenzhen, Guangdong 518107, P.R. China\\
\authormark{2}Department of Physics, University of Hong Kong, Pokfulam Road, Hong Kong SAR, China\\
\authormark{3}Institute of Quantum Information and Technology, Nanjing University of Posts and Telecommunications, Nanjing, 210023, P.R. China}

\email{\authormark{*}sunshh8@mail.sysu.edu.cn} 


\begin{abstract*} 
	Reference-frame-independent (RFI) quantum key distribution (QKD) presents promising advantages, especially for mobile-platform-based implementations, as it eliminates the need for active reference frame calibration. While RFI-QKD has been explored in various studies, limitations in key rate and distance persist due to finite data collection. In this study, we experimentally demonstrate an improved RFI-QKD protocol proposed by Zhu \textit{et al.} [Opt. Lett. 47, 4219 (2022)], featuring a statistical quantity for bounding information leaked to Eve that exhibits more insensitivity to statistical fluctuations and more robustness to variations in the reference frame. Taking into account finite-size considerations and potential general attacks, RFI-QKD is implemented over a distance of 175 \si{\kilo\meter} in this work. We believe that our study extends the communication distance achievable by RFI-QKD, thereby constituting a notable advancement for its practical application.
\end{abstract*}




\section{INTRODUCTION}

Quantum key distribution (QKD) facilitates the generation of an unconditionally secure key between two parties, Alice and Bob. Unlike the mathematical complexity of traditional cryptography, QKD achieves unconditional security through the fundamental principles of quantum mechanics, independent of any eavesdropper's capabilities. Following the inception of the first QKD protocol, BB84\cite{BB84}, numerous QKD protocols have been proposed to enhance performance in various situations, such as continuous variable(CV)-QKD \cite{CV}, passive round-robin differential phase-Shift(RRDPS)-QKD \cite{RRDPS}, device-independent(DI)-QKD \cite{acin_DIholevo_PRL2007}, measurement device independent(MDI)-QKD \cite{MDI, PanJW_MDI200km_PRL2014}, and others. The security of QKD under both collective and coherent attacks has been demonstrated with both perfect and imperfect devices\cite{XuFH_realisticdevice_2020RMP}. Additionally, QKD over long distances has been successfully implemented\cite{MDI_404km,TF_590km,PanJW_1000kmQKD_2023PRL,LiWei_110Mbps_2023np}, and several commercial QKD networks are available\cite{Shields_CambridgeQuantumnet_Nature2019}\cite{Qnet2}.

In most QKD protocols, a shared reference frame is vital to reduce quantum bit error rate (QBER) and enhance the key rate. However, practical variations such as temperature and atmospheric refraction necessitates active calibration or compensation, which potentially complicates the system or reduces the final key rate. To address this issue, Laing \textit{et al.} proposed reference-frame-independent (RFI) QKD\cite{RFI}.

In the standard BB84 QKD protocol, the estimation of phase error rate, proportional to $1-\cos(\beta)$ (where $\beta$ is the angle deviation of the reference frame between Alice and Bob), is essential to bound the information leaked to Eve. In RFI-QKD, a statistical quantity (\textit{\textbf{C}}), which is independent of $\beta$, is introduced instead of the phase error rate. RFI-QKD eliminates the need for compensation devices in reference frame calibration, which provides potential advantages in practical scenarios, such as mobile-platform-based QKD and fiber-free-space hybrid-channel-based QKD. Numerous experiments have been conducted to enhance RFI-QKD performance in practical situations, including RFI-MDI-QKD\cite{RFIMDI, RFIMDI2, ZhangCM_RFIMDIBIAS-JLT2017}, free-space RFI-QKD\cite{RFI4}, RFI-QKD with fewer states\cite{LiuHW_FewerStates_PRA2019,RFIfewer2}, and others.

However, a challenge in the practical implementation of RFI-QKD is that the finite data length introduces more fluctuation in statistical quantities, such as QBER and gain of the quantum state, worsening the key rate and maximal distance of RFI-QKD. To mitigate this problem, Zhu \textit{et al.}\cite{giveR} proposed an improved RFI-QKD protocol, introducing a new statistical quantity (\textit{\textbf{R}}) to bound information leakage. Compared with the statistical quantity \textit{C} in original RFI-QKD, the value of \textit{R} is more insensitive to statistical fluctuations of finite data length and more robust to variations in the reference frame. Thus, the key rate and maximal distance could be further increased.

In this work, we experimentally demonstrated the improved RFI-QKD protocol using different fiber channel links, achieving a maximal distance over 175 \si{\kilo\meter} with superconducting nanowires single photon detectors (SNSPD). Additionally, we show that, with commercial avalanche photodiode single photon detectors (APD), RFI-QKD could cover distances exceeding 100 \si{\kilo\meter} under a data length of $10^{11}$. Notably, we consider the security against general attacks for our experimental implement. We believe our findings mark a significant advancement in extending the practical application of RFI-QKD.

\section{PROTOCOL}
In contrast to BB84, the RFI-QKD protocol involves six quantum states in three bases (X, Y, Z). The Z-basis is well-aligned for generating secure keys, while the X- and Y-bases are used to estimate the information leaked to Eve. Due to environmental fluctuations, Bob's practical X- and Y-bases differ from those of Alice's. This distinction can be expressed as
\begin{align}
	X_B&=\cos \beta X_A+\sin \beta Y_A,\notag\\
	Y_B&=-\sin \beta X_A+\cos \beta Y_A,\\
	Z_A&=Z_B\notag, 
\end{align}
where subscripts A and B denote Alice and Bob, $\beta$ represents the deviation angle of the reference frame between them.

In the original RFI-QKD \cite{RFI}, a statistical quantity called \textbf{\textit{C}} is defined to estimate the leaked information, expressed by 
\begin{equation}
	C = \langle X_A X_B \rangle^2 +  \langle X_A Y_B \rangle^2 +  \langle Y_A X_B \rangle^2 + \langle Y_A Y_B \rangle^2. 
\end{equation}
It is straightforward to verify that \textit{C} is independent of $\beta$. Due to the unavoidable statistical fluctuations caused by finite sampled data, the final secret key rate decreases compared to that of infinite data. To mitigate this issue, various methods have been proposed. Recently, Zhu \textit{et al.} \cite{giveR} introduced a new protocol to address this challenge. In this improved RFI-QKD protocol, a new statistical quantity, denoted as \textit{\textbf{R}}, is introduced, defined by
\begin{equation}\label{R}
	R=\frac{1}{4}\left[\langle X_AX_B-Y_AY_B\rangle^2+
	\langle X_AY_B+Y_AX_B\rangle^2\right].
\end{equation}
Obviously, in comparison to \textit{C}, the new value \textit{R} exhibits lower sensitivity to statistical fluctuations. Further calculations reveal that the upper bound of information leaked to Eve, denoted as $I_E$, can be expressed as \cite{giveR}
\begin{equation}
	I_E\leq (1-e_{ZZ})h\left[\frac{1+\sqrt{R}-e_{ZZ}}{2(1-e_{ZZ})}\right]+e_{ZZ},
\end{equation}
where $h(x)$ is the binary Shannon entropy, and $e_{ZZ}$ is the bit error in the Z basis.

Additionally, due to the unavailability of commercial single-photon sources, a weak coherent source is commonly employed in most practical systems. Therefore, the decoy state method \cite{hwang_quantum_2003,ThreeDecoy,lo_decoy_2005} is always used to against the photon-number-splitting (PNS) attack \cite{PNS,Brassard_PracticalQKD_PRL2000}. In this context, we employ the standard ``weak + vacuum" decoy state method \cite{ma_practical_2005}, known for its effective performance with only two decoy states. It is important to note that the final key rate can be further enhanced by employing alternative decoy state methods, such as \cite{RFIonedecoy}. Subsequently, accounting for finite data length, the final secure key rate under a general attack is given by\cite{concisebound,Valerio_FiniteRotationRFI_NJP2010,ZhangCM_RFIMDIBIAS-JLT2017,LiuHW_FewerStates_PRA2019}
\begin{equation}\label{keyrate}
	\begin{split}
		r^L=&\frac{1}{N}[s^L_{zz,0}+s^L_{zz,1}(1-I_E^U)-s_{zz}fh(E^U_{zz})-
		\log_{2}{\frac{2}{\epsilon_{EC}}}-\\&2\log_{2}\frac{2}{\epsilon_{PA}}-7s_{zz}\sqrt{\frac{\log_{2}\frac{2}{\bar{\epsilon}}}{s_{zz}}}-30\log_{2}(N+1)].
	\end{split}
\end{equation}
Here, $s_{zz}$ and $E_{zz}$ represent the number of valid events and bit error rate, respectively, in which case Alice prepares a quantum state in the Z basis and Bob also measures it in the Z basis. $s_{zz,0}$ and $s_{zz,1}$ represent the number of vacuum events and single photon events, respectively, which should be estimated by the decoy state method. $\epsilon_{EC}(\epsilon_{PA})$ denotes the failure probability of error correction (privacy amplification), and $\bar{\epsilon}$ represents the accuracy of smooth min-entropy estimation  \cite{Valerio_FiniteRotationRFI_NJP2010}. For convenience, we set $\epsilon_{EC}=\epsilon_{PA}=\bar{\epsilon}=10^{-10}$. $N$ represents the length of the quantum state sent by Alice. The upper bound of \textit{\textbf{R}} is required to bound the information leaked to Eve ($I_E$). To estimate upper and lower bounds of single photon events for both X- and Y-bases, we use the decoy state method \cite{concisebound}, with introducing statistical fluctuation to measured parameters, we have
\begin{equation}
	s^U_{\chi\gamma,1}=\tau_1\frac{\frac{e^{\nu}}{p_{\nu}}N^U_{\chi\gamma,\nu}-\frac{e^{\omega}}{p_{\omega}}N^L_{\chi\gamma,\omega}}{\nu-\omega},
	\label{1upperbound}
\end{equation}
\begin{equation}
	\begin{split}
		&s^L_{\chi\gamma,1}=
		\frac{\tau_1\mu}{\mu(\nu-\omega)-\nu^2+\omega^2}\left[
		\frac{e^{\nu}}{p_{\nu}}N^L_{\chi\gamma,\nu}-\right.\\&\left.
		\frac{e^{\omega}}{p_{\omega}}N^U_{\chi\gamma,\omega}- \frac{\nu^2-\omega^2}{\mu^2}
		\left(\frac{e^{\mu}}{p_{\mu}}N^U_{\chi\gamma,\mu}-
		\frac{N^L_{\chi\gamma,0}}{\tau_0}\right) 
		\right],
	\end{split}
	\label{1lowerbound}
\end{equation}
and the lower bound of the vacuum events $s^L_{\chi\gamma,0}$ is given by

\begin{equation}
	s^L_{\chi\gamma,0}=
	\tau_0\frac{\nu\frac{e^{\omega}}{p_{\omega}}N^L_{\chi\gamma,\omega}-\omega\frac{e^{\nu}}{p_{\nu}}N^U_{\chi\gamma,\nu}}
	{\nu-\omega}.
	\label{0lowerbound}
\end{equation}
Here $k \in \{\mu, \nu, \omega\}$ is the intensity of decoy states. Without loss of generality, we assume that $\mu \gg \nu>\omega=0$. $\{p_\mu, p_\nu, p_\omega\}$ are the probabilities of corresponding intensities.  $\tau_{n}=\sum\limits_{k=\omega,\nu,\mu} p_k\frac{e^{-k}k^n}{n!}$ is the average probability that Alice prepares n-photon states. Based on Hoeffding's inequality \cite{Hoeffding}, the upper and lower bounds of measured parameters in experiment are given by
\begin{equation}
	\begin{split}
		N^U_{\chi\gamma,k}&=N_{\chi\gamma,k}+\sqrt{\frac{N_{\chi\gamma}}{2}\ln \frac{1}{\epsilon_{PE}}},\\
		N^L_{\chi\gamma,k}&=N_{\chi\gamma,k}-\sqrt{\frac{N_{\chi\gamma}}{2}\ln \frac{1}{\epsilon_{PE}}},
	\end{split}
\end{equation} 
where $N_{\chi\gamma}=\sum\limits_{k=\omega,\nu,\mu}N_{\chi\gamma,k}$, $\epsilon_{PE}$ is the failure probability of parameter estimation that the real value is out of the fluctuation range. Based on the composability of the security \cite{Valerio_FinitePractical_NJP2009}, the probability that our protocol fails is $\epsilon=\epsilon_{EC}+\epsilon_{PA}+\bar{\epsilon}+n_{PE}\epsilon_{PE}$. Here $n_{PE}$ is the number of parameters needed to be estimated. In our protocol,
through Eq.~[\ref{1upperbound}-\ref{0lowerbound}], $n_{PE}$ comes from three aspects: 1. two kinds of the counts that the number of the valid events
and the error events. 2. three decoy intensities $\{\mu,~\nu,~\omega\}$.
3. three kinds of events  $\{ ZZ,~ X_AX_B-Y_AY_B,~ X_AY_B+Y_AX_B\}$. In all, we get $n_{PE}=2\times 3\times 3=18$. Assuming $\epsilon_{PE}=10^{-10}$ is equal for all the parameters, the total failure probability of fluctuation estimation is $n_{PE}\epsilon_{PE}=1.8\times 10^{-9}$. Finally, the probability that our protocol fails is $\epsilon=\epsilon_{EC}+\epsilon_{PA}+\bar{\epsilon}+n_{PE}\epsilon_{PE}=2.1\times 10^{-9}$.

\section{EXPERIMENTAL SETUP AND RESULTS}

\begin{figure}[H]
	\centering
	\includegraphics[width=10cm]{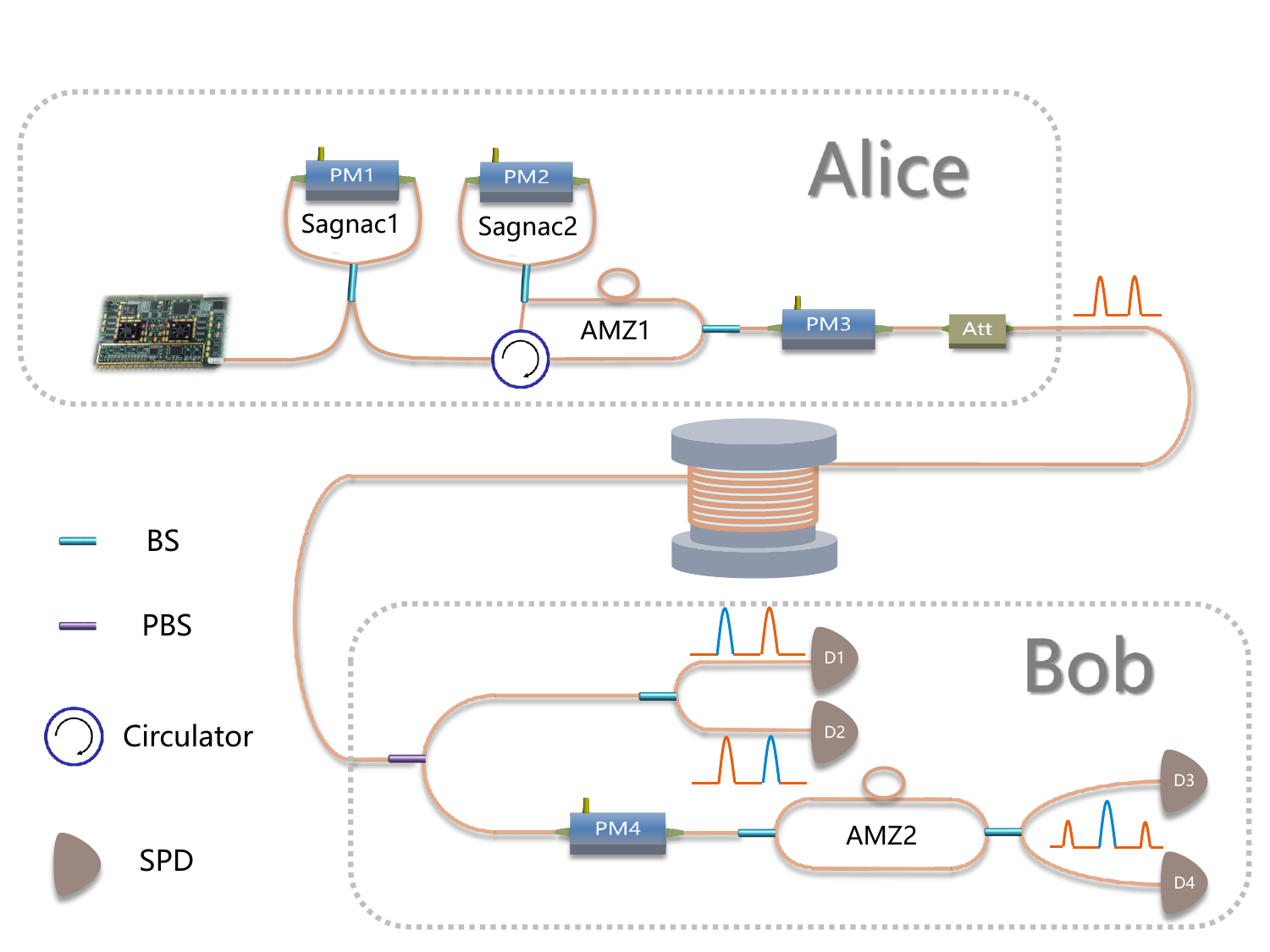}
	\caption{Experiment setup diagram. BS: Beam Splitter, PBS:  Polarization Beam Splitter, SPD: Single Photon Detector, PM:  Phase Modulation, Att: Optical Attenuator. The pulses are marked by orange of which the blue pulse indicates the detected pulse.}
	\label{experiment diagram}
\end{figure}

The experimental setup, illustrated in Fig.~\ref{experiment diagram}, is grounded in time-bin encoding. A homemade laser diode produces optical pulses with a width of approximately 50~\si{\pico\second} and a repetition rate of 100~\si{\mega\hertz}. Subsequently, the optical pulse traverses the first Sagnac Interferometer (Sagnac1) to generate different decoy states. The quantum state encoding module consists of the second Sagnac Interferometer (Sagnac2) and an asymmetric Mach–Zehnder interferometer (AMZ1).  Sagnac2 controls the path of the optical pulse in AMZ1. Modulating Sagnac2 with a phase of 0 or $\pi$ using a phase modulator (PM2) directs the optical pulse through the short or long path of AMZ1, generating quantum states $Z_0$ or $Z_1$. The time delay between $Z_0$ and $Z_1$ in our system is approximately 3~\si{\nano\second}. When a phase $\pi/2$ is modulated by PM2, a quantum state $| X_0\rangle = ( |Z_0\rangle + |Z_1\rangle)/\sqrt{2}$ is generated. Lastly, PM3 is employed to generate additional quantum states in the X and Y bases, namely, $X_1$, $Y_0$, and $Y_1$.

When the optical pulse traverses the quantum channel controlled by Eve and reaches Bob's region, Bob utilizes a polarization controller (not depicted in the figure) to randomize the polarization of the optical pulse. Subsequently, Bob passively selects either the Z- or XY- bases for measurement. In the Z-basis, a beam splitter (BS) and two single-photon detectors (D1 and D2) are employed to discriminate the time-bin of the optical pulse. On the other hand, for the X- and Y-bases, the quantum state is decoded using AMZ2 and PM4, followed by two other single-photon detectors (D3 and D4).

Simultaneously, a fully integrated custom-made electrical board is developed for both Alice and Bob to control the QKD system. This electrical board encompasses four modulators, an field programmable gate array (FPGA)-based processor, a four-channel radio frequency (RF) driver, a single-photon detector (SPD) controller module, and two lasers. The FPGA-based processor serves as the central component, responsible for generating, acquiring, and processing the digital signals. The four-channel RF driver, designed for phase modulation (PM), can produce an RF signal with an adjustable voltage ranging from 0 to 8V. The four-channel SPD controller generates trigger clocks for all SPDs and captures their output. The electrical board houses two lasers: one generates quantum optical pulses with a width of approximately 50~\si{\pico\second} and a wavelength of 1550~\si{\nano\metre}, while the other produces synchronized laser pulses with a wavelength of 1570~\si{\nano\metre} and a repetition rate of 10~\si{\mega\hertz}.

	\begin{table}[tbp]
		\caption{Parameters in simulations. $e_0$ is the intrinsic error rate of the system. $\alpha$ is the loss coefficient of fiber.  $\eta_Z$ and $\eta_{XY}$ are the loss of Z path and XY path in receiver, respectively. $e_d$ and $\eta_{det}$ are the dark count rate and efficiency of detectors, respectively. $f$ is the efficiency of error correction. Furthermore, $3~dB$ loss in valid pulse choosing due to time-bin detection is also been considered which isn't shown in the table.}
		\resizebox{\textwidth}{!}{
			\renewcommand\arraystretch{1.5}
			\begin{tabular}{cccccccccc}
				\hline \hline
				$\alpha$ \quad & \quad $\eta_Z$ \quad & \quad $\eta_{XY}$ \quad & \quad $f$ \quad & \quad $e_0$(APD) \quad & \quad $e_d$(APD) \quad & \quad $\eta_{det}(APD)$ \quad & \quad $e_0$(SNSPD) \quad & \quad $e_d$(SNSPD) \quad & \quad $\eta_{det}$(SNSPD) \quad
				\\ \hline
				$0.2\ dB/km$ \quad & \quad$10\ dB$ \quad & \quad $12\ dB$ \quad & \quad 1.16 \quad & \quad 1\% \quad & \quad $8\times 10^{-6}$ \quad & \quad $15\%$ \quad &\quad 0.2\% \quad & \quad $10^{-8}$ \quad & \quad 55\%
				\\ \hline		\hline
		\end{tabular}}
		\label{parasetup}	
	\end{table}

\begin{figure}[H] 
	\centering
	\includegraphics[width=0.5\columnwidth]{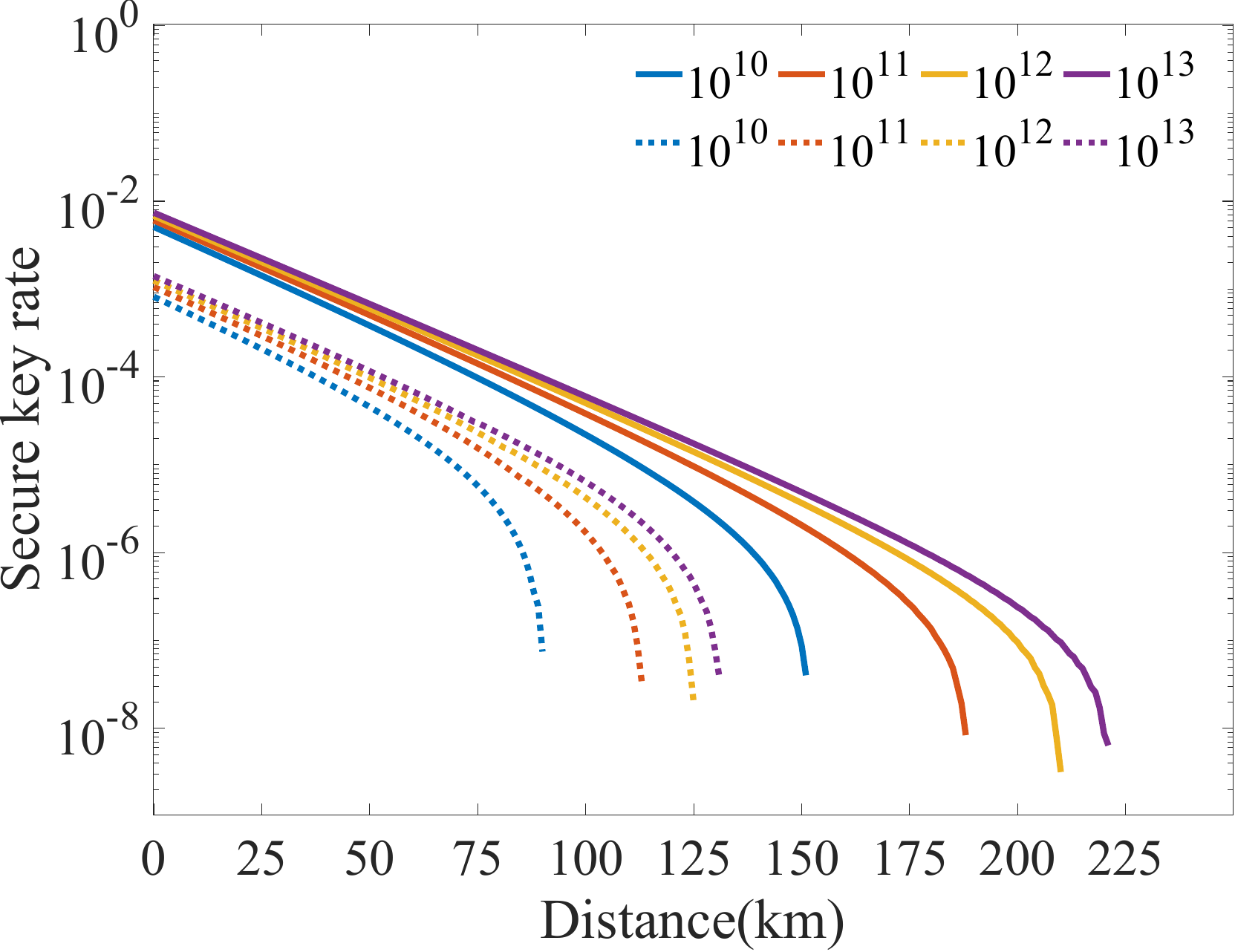}		
	\newline
	\caption{Simulation results. The solid lines are key rate by adopting SNSPD, and the dotted lines are key rate by adopting our homemade APD detector. The lines from left to right represent the key rate with $N = \{10^{10},10^{11},10^{12},10^{13}\}$ respectively for both SNSPD and APD.}
	\label{Simulation_results}	
\end{figure}

To assess our system's performance, we initially conduct simulations to determine the key rate, optimizing parameters ${\mu, \nu, \omega, p_\mu, p_\nu, p_\omega, p_Z}$ individually for each distance. Here, $p_Z$ represents the probability of the Z-basis for both Alice and Bob, while the probabilities for the X- and Y- bases are identical, denoted as $p_X=p_Y=(1-p_Z)/2$. The remaining parameters required for simulations are obtained directly from measurements performed with our experimental setup, as detailed in TABLE.~\ref{parasetup}. The simulation results, presented in Fig.~\ref{Simulation_results}, unambiguously indicate the system's feasibility for achieving a maximal distance exceeding 200 kilometers in our enhanced RFI-QKD system. Furthermore, as noted earlier, leveraging alternative decoy state methods and more efficient finite data analysis theories holds the potential for further enhancement of the final key rate.

\begin{figure}[H]
	\centering
	\includegraphics[width=0.5\columnwidth]{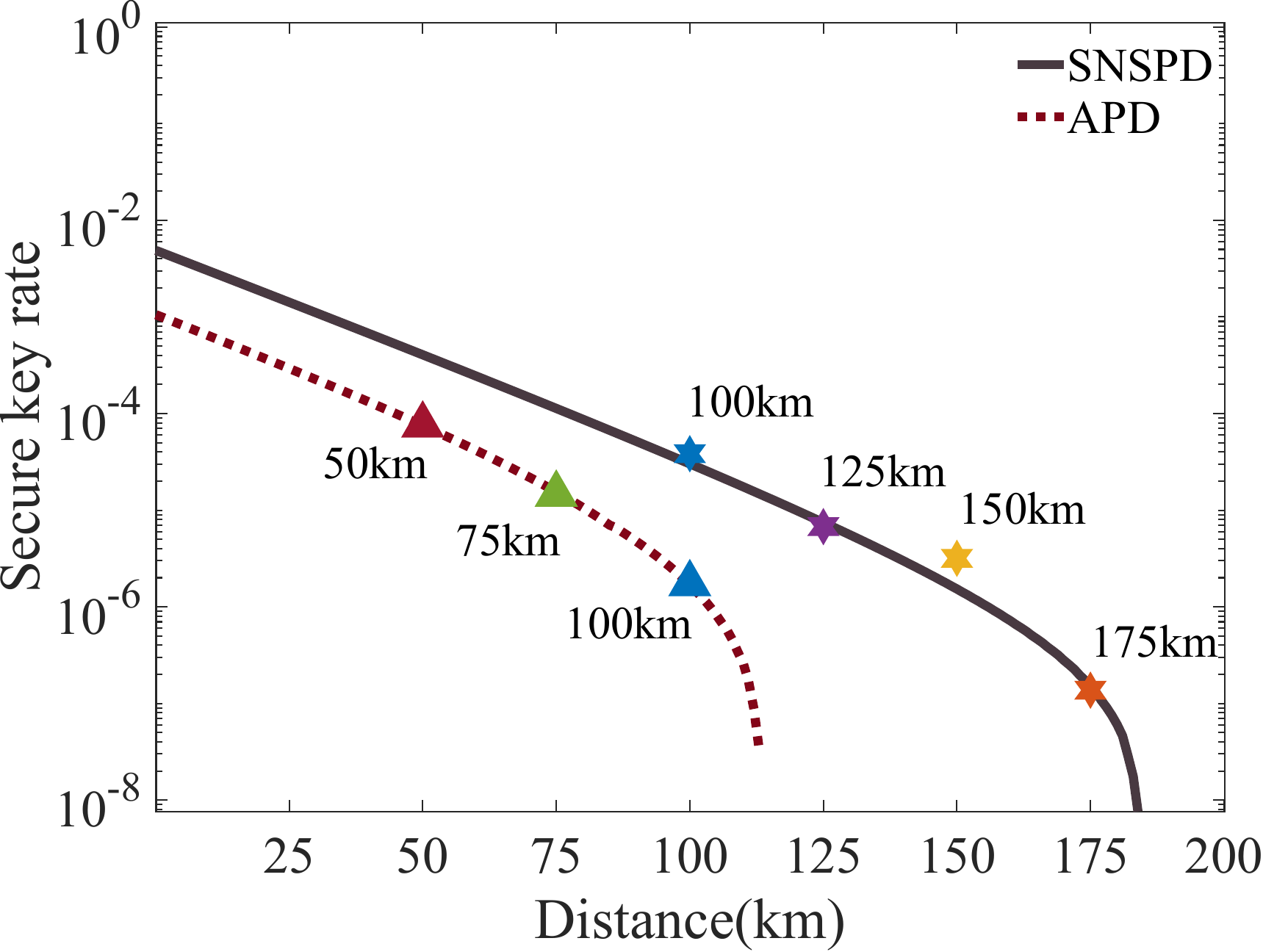}
	\newline	
	\caption{Experimental results. The triangles are measured key rate with APD at 50, 75 and 100~\si{\kilo\metre}. The hexagrams are measured key rate with SNSPD at 100~\si{\kilo\metre}, 125~\si{\kilo\metre}, 150~\si{\kilo\metre} and 175~\si{\kilo\metre}. The total data length in our experiment is $N=10^{11}$. The two curves are the simulation results with the experimental parameters.}
	\label{Experimental_results}
\end{figure}

While a longer data length $N$ generally yields higher key rate, practical considerations, such as the slow rotation of the reference frame and the 100MHz repetition rate in our system. We set $N=10^{11}$ based on simulation outcomes and our experimental configuration. Consequently, the acquisition time for each round of key change is approximately 1000~\si{\second}. Subsequently, experiments were conducted using both APD and SNSPD. For the APD-based experiments, fiber lengths of 50, 75 and 100~\si{\kilo\metre} were considered. Simultaneously, we enhanced system performance by employing SNSPD at distances of 100, 125, 150 and 175~\si{\kilo\metre}. The final key rate is shown in Fig.~\ref{Experimental_results}, and other detailed results are listed in Table~\ref{Result_table}. Notably, all experimental results align closely with the simulation predictions.

\begin{table*}[htbp]
	\caption{Experiment result. $r^L$ and $R^L$ are the lower bounds of secure key rate and $R$, respectively. $E_{ZZ,1}^U$ is the upper bound of the error rate and $Q_{ZZ,1}^L$ is the lower bound of the single events in ZZ bases. $E_{ZZ}$ and $Q_{ZZ}$ are the error rate and the number of valid events of ZZ bases in the experiment.}
	\resizebox{\textwidth}{!}{
		\renewcommand\arraystretch{1.5}
		\begin{tabular}{ccccccccccccccc}
			\hline \hline
			Detector  \quad & \quad $ Distance $ \quad & \quad $ r^L $ \quad & \quad $ R^L $ \quad & \quad $ E_{ZZ,1}^U $ \quad & \quad $ Q_{ZZ,1}^L $ \quad & \quad $ E_{ZZ} $ \quad & \quad $ Q_{ZZ} $ \quad & \quad $ P_Z $ \quad & \quad $ \mu $ \quad & \quad $ \nu $ \quad & \quad $ \omega $ \quad & \quad $ P_\mu $ \quad & \quad $ P_\nu $ \quad & \quad  $P_\omega$ 
			\\ \hline
			APD  \quad & \quad $ 50 $ \quad & \quad $ 7.69\times 10^{-5} $ \quad & \quad $ 0.59 $ \quad & \quad $ 1.9\% $ \quad & \quad $ 2.26\times 10^{7} $ \quad & \quad $ 1.40\% $ \quad & \quad $ 3.06\times 10^{7} $ \quad & \quad $ 0.90 $ \quad & \quad $ 0.58 $ \quad & \quad $ 0.27 $ \quad & \quad $ 0 $ \quad & \quad $ 0.52 $ \quad & \quad $ 0.36 $ \quad & \quad  0.11 \\ \hline
			APD  \quad & \quad $ 75 $ \quad & \quad $ 1.50\times 10^{-5} $ \quad & \quad $ 0.64 $ \quad & \quad $ 3.0\% $ \quad & \quad $ 5.44\times 10^{6} $ \quad & \quad $ 2.15\% $ \quad & \quad $ 8.64\times 10^{6} $ \quad & \quad $ 0.86 $ \quad & \quad $ 0.58 $ \quad & \quad $ 0.28 $ \quad & \quad $ 0 $ \quad & \quad $ 0.36 $ \quad & \quad $ 0.48 $ \quad & \quad  0.16 \\ \hline
			APD  \quad & \quad $ 100 $ \quad & \quad $ 2.24\times 10^{-6} $ \quad & \quad $ 0.78 $ \quad & \quad $ 5.8\% $ \quad & \quad $ 1.26\times 10^{6} $ \quad & \quad $ 3.90\% $ \quad & \quad $ 2.68\times 10^{6} $ \quad & \quad $ 0.74 $ \quad & \quad $ 0.56 $ \quad & \quad $ 0.28 $ \quad & \quad $ 0 $ \quad & \quad $ 0.30 $ \quad & \quad $ 0.46 $ \quad & \quad  0.25 \\ \hline
			SNSPD  \quad & \quad $ 100 $ \quad & \quad $ 4.05\times 10^{-5} $ \quad & \quad $ 0.66 $ \quad & \quad $ 0.45\% $ \quad & \quad $ 9.04\times 10^{6} $ \quad & \quad $ 0.47\% $ \quad & \quad $ 1.70\times 10^{7} $ \quad & \quad $ 0.90 $ \quad & \quad $ 0.58 $ \quad & \quad $ 0.26 $ \quad & \quad $ 0 $ \quad & \quad $ 0.46 $ \quad & \quad $ 0.41 $ \quad & \quad  0.13 \\ \hline
			SNSPD  \quad & \quad $ 125 $ \quad & \quad $ 7.59\times 10^{-6} $ \quad & \quad $ 0.74 $ \quad & \quad $ 0.57\% $ \quad & \quad $ 1.40\times 10^{6} $ \quad & \quad $ 0.37\% $ \quad & \quad $ 2.07\times 10^{6} $ \quad & \quad $ 0.87 $ \quad & \quad $ 0.59 $ \quad & \quad $ 0.26 $ \quad & \quad $ 0 $ \quad & \quad $ 0.35 $ \quad & \quad $ 0.48 $ \quad & \quad  0.17 \\ \hline
			SNSPD  \quad & \quad $ 150 $ \quad & \quad $ 2.97\times 10^{-6} $ \quad & \quad $ 0.86 $ \quad & \quad $ 2.3\% $ \quad & \quad $ 5.08\times 10^{5} $ \quad & \quad $ 2.0\% $ \quad & \quad $ 5.20\times 10^{5} $ \quad & \quad $ 0.83 $ \quad & \quad $ 0.59 $ \quad & \quad $ 0.27 $ \quad & \quad $ 0 $ \quad & \quad $ 0.33 $ \quad & \quad $ 0.46 $ \quad & \quad  0.21 \\ \hline
			SNSPD  \quad & \quad $ 175 $ \quad & \quad $ 1.64\times 10^{-7} $ \quad & \quad $ 0.14 $ \quad & \quad $ 1.04\% $ \quad & \quad $ 3.42\times 10^{5} $ \quad & \quad $ 0.950\% $ \quad & \quad $ 3.00\times 10^{5} $ \quad & \quad $ 0.69 $ \quad & \quad $ 0.59 $ \quad & \quad $ 0.23 $ \quad & \quad $ 0 $ \quad & \quad $ 0.31 $ \quad & \quad $ 0.45 $ \quad & \quad  0.24 \\ 
			\hline\hline
	\end{tabular}}
	\label{Result_table}	
\end{table*}

\section{CONCLUSION}
RFI-QKD exhibits potential advantages in practical applications, as it eliminates the need for active reference frame calibration. Various efforts have been dedicated to enhancing RFI-QKD's performance and addressing practical challenges. In this paper, we present an experimental demonstration of an improved RFI-QKD protocol, which is designed to be less susceptible to statistical fluctuations in finite sampled data. This enhancement promises increasing key rates and extending the maximal distance achievable by RFI-QKD. With finite-size issues considered, we conduct experiments with different fiber distances, achieve a remarkable maximal transmission distance exceeding 175~\si{\kilo\metre}. Consequently, our work constitutes a substantial advancement for the practical implementation of RFI-QKD.

\begin{backmatter}
	\bmsection{Funding}
	This work was supported by Shenzhen Science and Technology Program (Grant No. JCYJ20220818102014029); National Natural Science Foundation of China (Grants No. 62171458 and No. 62371244). 
	\bmsection{Disclosures}
	 The authors declare no conflicts of interest.
	 \bmsection{Data availability} 
	 Data underlying the results presented in this paper are not publicly available at this time but may be obtained from the authors upon reasonable request
	 
\end{backmatter}

\appendix

\section{CHANNEL MODEL}
\setcounter{equation}{0}
\renewcommand\theequation{A\arabic{equation}}
We employ the model that is proposed in \cite{RFIsourceflaw}\cite{RFIsourceflawexp}. Denoting the conditional probability for single photon that Alice sends the state $\ket{\chi_i}$ and Bob obtains $\ket{\gamma_j}$ as $V_{\chi_i|\gamma_j}$. Then, the gain in $k$ intensity with WCS source by can be calculated by
\begin{equation}
	\begin{split}
		&V_{k,\chi_i|\gamma_j}=\sum_{n=0}^{\infty} Y_n\frac{k^n}{n!}e^{-k}
		=\frac{1}{2}\!+
		\!\frac{1\!-\!e_d}{2}\\
		&[(1\!+\!\eta C_{\chi_i\!|\gamma_j}\!-\!\eta)^n\!-\!(1\!-\!\eta C_{\chi_i\!|\!\gamma_j}\!)^n\!-\!(1\!-\!e_d)(1\!-\!\eta)^n],
	\end{split}
\end{equation}		
where $\eta$ is the transmittance of the total system. $ C_{\chi_i|\gamma_j}=|\braket{\chi_i|\gamma_j}|^2$ is the theoretical probability that Bob obtains $\ket{\gamma_j}$ while Alice sends the state $\ket{\chi_i}$. We suppose using two detectors and their dark counts are both $e_d$. The first term denotes the correct click generated by signal photon, the second term is the click caused by dark count, the third term represents randomly assigning a detector click when two detectors simultaneously click. 

Then we can calculate single photon gain $Q_{\chi\gamma}$ and QBER in $\chi\gamma$ basis by 
\begin{equation}
	\begin{split}
		Q_{\chi\gamma}=\frac{1}{2}(V_{\chi_0|\gamma_0}&+V_{\chi_0|\gamma_1}+V_{\chi_1|\gamma_0}+V_{\chi_1|\gamma_1}),\\
		E'_{\chi\gamma}=&\frac{V_{\chi_0|\gamma_1}+V_{\chi_1|\gamma_0}}{2Q^1_{\chi_0|\gamma_0}}.
	\end{split}
	\label{channel model}
\end{equation}

Considering the bit flip probability caused by intrinsic optical system error which we denote as $e_0$, the QBER can be rewritten as 
\begin{equation}
	E_{\chi\gamma}=e_0(1-E'_{\chi\gamma})+E'_{\chi\gamma}.
	\label{E}
\end{equation}

In addition, finite-size case with decoy-state method are considered. Hence, the numbers of the corresponding events that  after considering the bases and decoy states choosing probability are
\begin{equation}
	\begin{split}
		N_{\chi\gamma,k}=NPr_{\chi}^{A}Pr_{\gamma}^B Q_{\chi\gamma,k} Pr_k,
		\\
		m_{\chi\gamma,k}=NPr_{\chi}^{A}Pr_{\gamma}^B E_{\chi\gamma,k} Pr_k.
	\end{split}
	\label{nm}
\end{equation} 

\section{Finite-size issue}
\setcounter{equation}{0}
\renewcommand\theequation{B\arabic{equation}}
In this appendix, we state the details of the estimation calculation of the events number in the finite case. First, we derive the analytical bounds of the vacuum events and single photon events in asymptotic case.

For lower-bound of the number of vacuum events, using the formulation proposed by Ref.\cite{concisebound}, and the counts in different bases calculated by \ref{nm}, we have
\begin{equation}
	\begin{split}
		&\nu\frac{e^{\omega}}{p_{\omega}}N_{\chi\gamma,\omega}-\omega\frac{e^{\nu}}{p_{\nu}}N_{\chi\gamma,\nu}\\
		&=\frac{(\nu-\omega)N_{\chi\gamma,0}}{\tau_0}-
		\nu\omega\sum_{n=2}^{\infty} \frac{(\nu^{n-1}-\omega^{n-1})N_{\chi\gamma,n}}{n!\tau_n},
	\end{split}
\end{equation}
where $\tau_{n}=\sum\limits_{k=\omega,\nu,\mu} p_k\frac{e^{-k}k^n}{n!}$ is the probability that Alice prepares n-photon state.

Then the number of vacuum events has the relation that
\begin{equation}
	N_{\chi\gamma,0}\geq
	\tau_0\frac{\nu\frac{e^{\omega}}{p_{\omega}}N_{\chi\gamma,\omega}-\omega\frac{e^{\nu}}{p_{\nu}}N_{\chi\gamma,\nu}}
	{\nu-\omega}=:N^L_{\chi\gamma,0}.
	\label{0lowbound}
\end{equation} 

For single-photon events, we have
\begin{equation}
	\begin{split}
		&\frac{e^{\omega}}{p_{\omega}}N_{\chi\gamma,\omega}-\frac{e^{\nu}}{p_{\nu}}N_{\chi\gamma,\nu}\\
		&=\frac{(\nu-\omega)N_{\chi\gamma,1}}{\tau_1}+\sum_{n=2}^{\infty}\frac{(\nu^n-\omega^n)N_{\chi\gamma,n}}{n!\tau_{n}}\\
		&\leq\frac{(\nu-\omega)N_{\chi\gamma,1}}{\tau_1}+\frac{\nu^2-\omega^2}{\mu^2}\sum_{n=2}^{\infty}\frac{\mu^n N_{\chi\gamma,n}}{N_{\chi\gamma,1}}\\
		&\leq\frac{(\nu-\omega)N_{\chi\gamma,1}}{\tau_1}+\frac{\nu^2-\omega^2}{\mu^2}(\frac{e^\mu N_{\chi\gamma,\mu}}{p_{\mu}}-\frac{N_{\chi\gamma,0}}{\tau_0}-\frac{\mu N_{\chi\gamma,1}}{\tau_1})
	\end{split},
\end{equation}
where we use the multiphoton events
\begin{equation*}
	\sum_{n=2}^{\infty}\frac{\mu^n N_{\chi\gamma,n}}{N_{\chi\gamma,1}}=\frac{\nu^2-\omega^2}{\mu^2}(\frac{e^\mu N_{\chi\gamma,\mu}}{p_{\mu}}-\frac{N_{\chi\gamma,0}}{\tau_0}-\frac{\mu N_{\chi\gamma,1}}{\tau_1}).
\end{equation*}
Then the lower bound of single-photon events is
\begin{small}
	\begin{equation}
		\begin{split}
			N_{\chi\gamma,1}&\geq \frac{\tau_1\mu}{\mu(\nu-\omega)-\nu^2+\omega^2}\\&\left[
			\frac{e^{\nu}}{p_{\nu}}N_{\chi\gamma,\nu}-\right.\left.
			\frac{e^{\omega}}{p_{\omega}}N_{\chi\gamma,\omega}- \frac{\nu^2-\omega^2}{\mu^2}
			\left(\frac{e^{\mu}}{p_{\mu}}N_{\chi\gamma,\mu}-
			\frac{N_{\chi\gamma,0}}{\tau_0}\right) 
			\right]\\
			&=:N^L_{\chi\gamma,1},
		\end{split}
		\label{1lowbound}
	\end{equation}
\end{small}

Then, we need the upper bound of the single-photon events, which can be obtained by 
\begin{equation}
	N_{\chi\gamma,1}=\tau_1\frac{\frac{e^{\nu}}{p_{\nu}}N_{\chi\gamma,\nu}-\frac{e^{\omega}}{p_{\omega}}N_{\chi\gamma,\omega}}{\nu-\omega}=:N^U_{\chi\gamma,1}.
	\label{1upbound}
\end{equation}

By introducing in the statistic influence, which we denote as
\begin{eqnarray}
	N^*\leq N+\delta(N,\epsilon_{PE}),\\
	N^*\geq N-\delta(N,\epsilon_{PE}).
\end{eqnarray}
Here $N$ is the observed quantity and $N^*$ represents the standard value. Using Hoeffding's inequality for independent events, with the probability at least $1-2\epsilon_{PE}$, the fluctuation quantity has  $\delta(N,\epsilon_{PE})=\sqrt{\frac{N_{\chi\gamma,k}}{2}\ln \frac{1}{\epsilon_{PE}}}$.

Finally, by replacing the quantity with fluctuation in Eq.~[\ref{0lowbound}-\ref{1upbound}], we get the Eq.~[\ref{1upperbound}-\ref{0lowerbound}].
\bibliography{improvedbib}





\end{document}